\documentclass[12pt]{article}
\usepackage{graphicx}
\usepackage{setspace}
\usepackage{cite}
\usepackage{color}

\newcommand\pubnumber{}
\newcommand\pubdate{July 6, 2015}

\def\Title#1{\begin{center} {\Large #1 } \end{center}}
\def\Author#1{\begin{center}{ \sc #1} \end{center}}
\def\Address#1{\begin{center}{ \it #1} \end{center}}

\newcommand\pubblock{\rightline{\begin{tabular}{l} \pubnumber \\
    \pubdate  \end{tabular}}}
\newenvironment{Abstract}{\begin{quotation}  }{\end{quotation}}
\newenvironment{Presented}{\begin{quotation} \begin{center} 
    PRESENTED AT\end{center}\bigskip 
    \begin{center}\begin{large}}{\end{large}\end{center} \end{quotation}}

\def\beq{\begin{equation}}
\def\eeq#1{\label{#1}\end{equation}}
\def\eeqn{\end{equation}}

\def\beqa{\begin{eqnarray}}
\def\eeqa#1{\label{#1}\end{eqnarray}}
\def\eeqan{\end{eqnarray}}

\let\bar=\overbar

\def\Dslash{\not{\hbox{\kern-4pt $D$}}}
\def\dslash{\not{\hbox{\kern-2pt $\del$}}}

\def\msb{{\bar{\ssstyle M \kern -1pt S}}}

\begin{document}
\begin{titlepage}
\pubblock

\vfill
\begin{spacing}{1.5}
\Title{A Study of Active Shielding Optimized for \\
1--80 keV Wide-Band X-ray Detector in Space}
\end{spacing}
\vfill
\Author{Yoshihiro Furuta$^1$, Yuki Murota$^1$, Junko S. Hiraga$^2$, \\
Makoto Sasano$^1$, Hiroaki Murakami$^1$, and Kazuhiro Nakazawa$^1$}
\Address{$^1$ Department of Physics, Graduate School of Science, \\
The University of Tokyo, 7-3-1 Hongo, Bunkyo-ku, Tokyo 113-0033, Japan \\
$^2$ Department of Physics, School of Science and Technology, \\
Kwansei Gakuin University, 2-1 Gakuen, Sanda, Hyogo 669-1337, Japan}
\vfill
\begin{Abstract}
Active shielding is an effective technique to reduce background signals in hard X-ray detectors and to enable observing darker sources with high sensitivity in space. Usually the main detector is covered with some shield detectors made of scintillator crystals such as BGO (Bi$_4$Ge$_3$O$_{12}$), and the background signals are filtered out using anti-coincidence among them. Japanese X-ray observing satellites ``\textit{Suzaku}" and ``\textit{ASTRO-H}" employed this technique in their hard X-ray instruments observing at $>10$~keV.

In the next generation X-ray satellites, such as the NGHXT proposal, a single hybrid detector is expected to cover both soft (1--10~keV) and hard ($>10$~keV) X-rays for effectiveness. However, present active shielding is not optimized for the soft X-ray band, 1--10~keV. For example, Bi and Ge, which are contained in BGO, have their fluorescence emission lines around 10~keV. These lines appear in the background spectra obtained by \textit{ASTRO-H} Hard X-ray Imager, which are non-negligible in its observation energy band of 5--80~keV. 

We are now optimizing the design of active shields for both soft and hard X-rays at the same time. As a first step, we utilized a BGO crystal as a default material, and measured the L lines of Bi and K lines of Ge from it using the X-ray SOIPIX, ``XRPIX".
\end{Abstract}
\vfill
\begin{spacing}{1.2}
\begin{Presented}
International Workshop on SOI Pixel \\
Detector (SOIPIX2015), Tohoku University, \\
Sendai, Miyagi, Japan, June 3--6, 2015
\end{Presented}
\end{spacing}
\vfill
\end{titlepage}
\def\thefootnote{\fnsymbol{footnote}}
\setcounter{footnote}{0}

\section{Introduction}
\subsection{Active Shielding Used in X-ray Detectors}
X-ray observation of celestial objects tells us much about the high-energy phenomena in the universe. In the hard X-ray band ($>10$~keV), which is a sensitivity frontier, reduction of detector-originated background is crucial for accurate measurements, because the number of photons from the objects gets smaller.

\begin{figure}[htb]
\centering
\includegraphics[width=5in, clip]{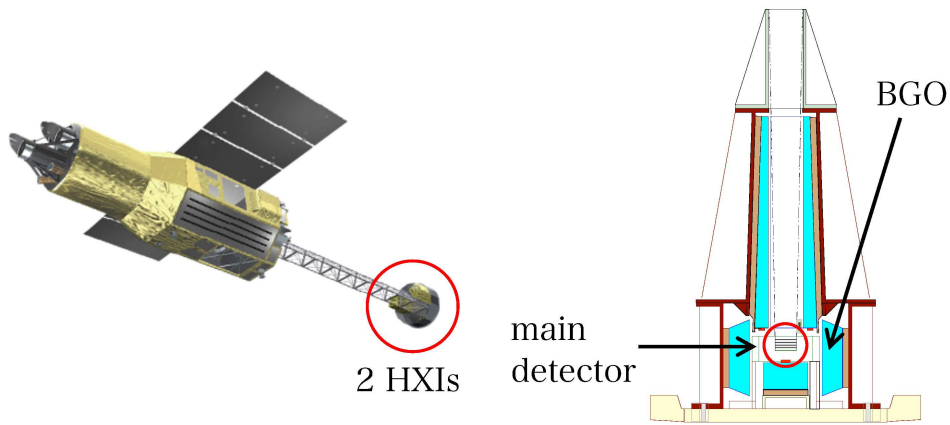}
\caption{(Left) Whole view of the \textit{ASTRO-H} satellite to be launched at the end of Japanese fiscal year 2015. At the tail of the satellite, marked by a red line, are the two HXIs (Hard X-ray Imagers). (Right) Cross section view of the HXI, where the main detector is marked by a red line and the BGO shields are colored in blue.}
\label{HXI}
\end{figure}

Active shielding is a technique widely used to reduce backgrounds significantly in hard X-ray detectors. The main detector is covered with another larger one, and cosmic-ray originated and scattered photon events are excluded using anti-coincidence between them. Japanese X-ray observing satellites ``\textit{Suzaku}" and ``\textit{ASTRO-H}" employed this technique in their hard X-ray instruments~\cite{suzaku_paper, ASTRO-H_paper, HXI_paper}. Figure \ref{HXI} shows \textit{ASTRO-H} and the Hard X-ray Imager (HXI) onboard it, which adopts BGO (Bi$_4$Ge$_3$O$_{12}$) scintillators as its active shields.

\subsection{Issues of Active Shielding at the Soft X-ray Band}

\begin{figure}[htb]
\centering
\includegraphics[width=4in, clip]{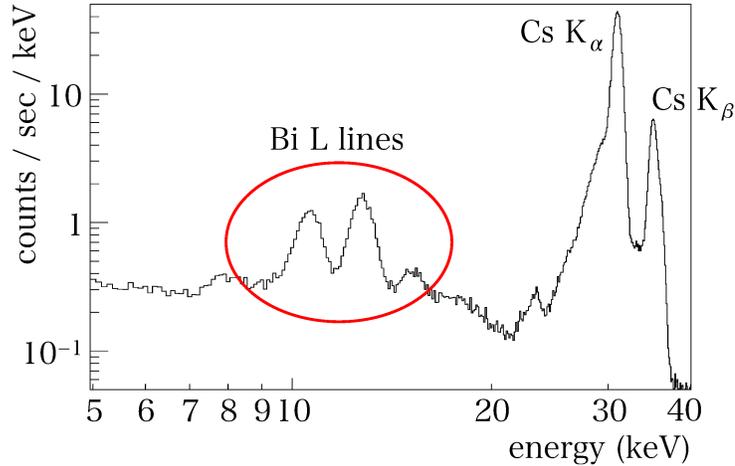}
\caption{X-ray spectrum of \textit{ASTRO-H} HXI DSSD (Double-sided Silicon Strip Detector) 1st layer after event selections obtained in the low temperature test using a $^{133}$Ba X-ray source. The three lines marked in red are the secondary X-rays from the BGO shields.}
\label{HXI_spectrum}
\end{figure}

BGO is one of the most suitable materials for active shielding. It has a large photo-absorption cross section because of the high atomic number of Bi ($Z=83$) and its high density ($\rho =7.13$~g/cm$^{-3}$). Additionally, it is not deliquescent and is easy to get large ingots. There is a possibility, however, that the characteristic X-ray fluorescent lines of Bi and Ge around 10~keV become nuisances in the soft X-ray (1--10~keV) spectra. In fact, as shown in Figure \ref{HXI_spectrum}, some secondary X-ray lines from the BGO shields were detected in the calibration tests of HXI on ground~\cite{LowTemp}. The same lines will also appear in space because of BGO excitation by cosmic X-ray background and BGO activation by high energy protons and neutrons, and become non-negligible as backgrounds in this band. In case of other familiar scintillators such as CsI or NaI, the same issues should occur because Cs and I have their fluorescence emission lines at $\sim 4$~keV and $\sim 30$~keV.

\section{Basic Studies for Wide-Band Application: \\ NGHXT}
NGHXT (Next Generation Hard X-ray Telescope) is a candidate future Japanese X-ray mission proposed as a next step to \textit{ASTRO-H}. The main aim is to obtain 1--80~keV wide-band high-angular/energy-resolution images and spectra with low background using active shielding.

Applying active shielding to both the hard ($>10$~keV) and the soft (1--10~keV) band at the same time is a new approach. Good selection of materials, combination of active and passive shields, and optimization between these choices are critically important. As a first step for this new attempt, we studied the X-ray lines from the BGO shield at the soft X-ray band using X-ray SOIPIX, or ``XRPIX", which was developed by Kyoto University and KEK for X-ray astronomical use, as our main detector.

\section{Experimental Setup}
\begin{figure}[htb]
\centering
\includegraphics[width=3in, clip]{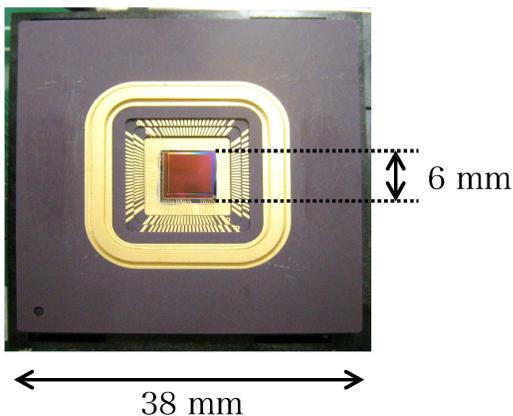}
\caption{Top view of XRPIX2b NFZ.}
\label{XRPIX2b}
\end{figure}

\begin{figure}[htb]
\centering
\includegraphics[width=4.5in, clip]{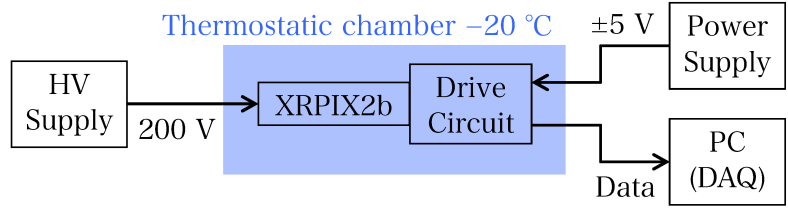}
\caption{Schematic view of the whole experimental setup. XRPIX2b and its drive circuit were put in the thermostatic chamber, which is colored in blue.}
\label{schematic}
\end{figure}

\begin{figure}[htb]
\centering
\includegraphics[width=3in, clip]{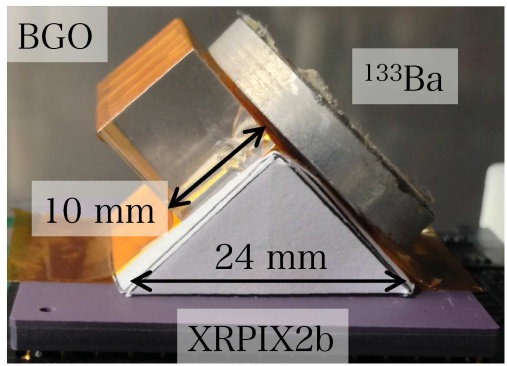}
\caption{Setup of the XRPIX2b, the BGO crystal, and the $^{133}$Ba X-ray source.}
\label{jig}
\end{figure}

We used the 4th XRPIX, ``XRPIX2b NFZ (N-type Floating Zone wafer)", which is shown in Figure \ref{XRPIX2b}. The active area is $4.32 \times 4.32$~mm$^2$, and the full depletion thickness is 500~$\mu$m~\cite{TD}. There are $144 \times 144$ pixels in the sensor and the pixel size is $30 \times 30$~$\mu$m$^2$ each.
The experimental setup is drawn in Figure \ref{schematic}. Whole setup was cooled to $-20$~$^\circ$C in the thermostatic chamber and the XRPIX2b was applied a reverse bias voltage of 200~V. The data were acquired by XRPIX2b DAQ software frame mode ver.16 developed by Kyoto University. As shown in Figure \ref{jig}, a BGO crystal of $10 \times 10 \times 20$~mm$^3$ and a $^{133}$Ba X-ray source of $\sim 1.1$~MBq were put on the triangular jig. The integration time of the DAQ frame mode was set to 2~msec/frame. Under this condition, we measured X-ray spectra of $^{133}$Ba with and without the BGO crystal, acquiring 50000 frames each so that the effective exposure was 100~sec. 

\section{Results}
\begin{figure}[htb]
\centering
\includegraphics[width=5.5in, clip]{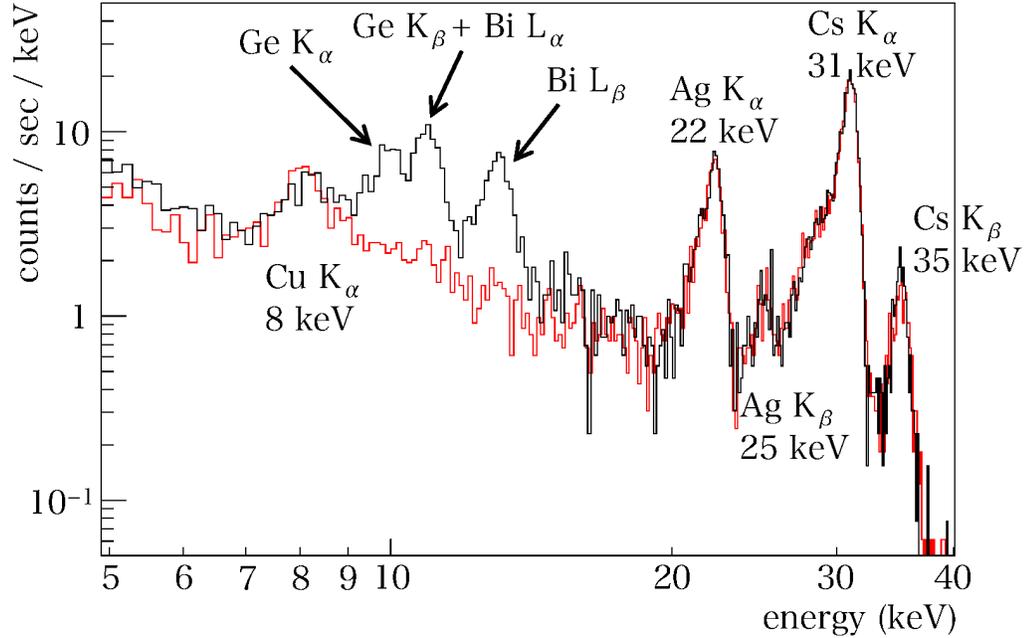}
\caption{Single-pixel event spectra of $^{133}$Ba with a BGO crystal (black) and $^{133}$Ba only (red) obtained by XRPIX2b using frame mode. Multi-pixel events are not included.}
\label{spectra}
\end{figure}

The black/red line in Figure \ref{spectra} shows the X-ray spectrum of $^{133}$Ba with/without the BGO crystal, respectively. In both spectra, K lines from $^{133}$Cs, which is the daughter nucleus of $^{133}$Ba, were detected. The energy resolution is $\sim 0.5$~keV (FWHM) at 31~keV.

In the black spectrum, we clearly see three lines around 10~keV which are not seen in the red one. These lines are the secondary X-ray lines from the BGO crystal, which are a Ge K$_\alpha$ line at 9.9~keV, a composite line of Ge K$_\beta$ at 10.8~keV and Bi L$_\alpha$ at 11.0~keV, and a Bi L$_\beta$ line at 13.0~keV.

We also detected K lines from Ag and a K line from Cu. These are considered to be the characteristic X-ray lines from surrounding materials (e.g. Ag paste for XRPIX packaging).

\section{Conclusion and Future Prospects}
We measured the secondary X-rays from BGO utilizing XRPIX in this study. In order to develop the optimized active shielding for NGHXT wide-band detector, we will construct anti-coincidence system using event-driven XRPIX and a BGO active shield in the next step, and then try to reduce the secondary X-ray signals by detecting coincident energy deposits in the active shield. We will also study about combination with thin passive shields such as ``Graded-Z" shield in the future.

\end{document}